\documentclass[twocolumn,aps,prb,superscriptaddress,showpacs,floatfix]{revtex4-1}
\usepackage[utf8]{inputenc}
\usepackage{color}
\usepackage{amsmath}
\usepackage{amssymb}
\usepackage{graphicx}
\usepackage[normalem]{ulem}
\usepackage[T1]{fontenc}
\usepackage[unicode=true,pdfusetitle,
bookmarks=true,bookmarksnumbered=false,bookmarksopen=false,
breaklinks=false,pdfborder={0 0 1},backref=false,colorlinks=false]
{hyperref}
\hypersetup{
	colorlinks,linkcolor=blue,citecolor=blue,urlcolor=blue}
\usepackage{float}
\usepackage{verbatim}
\usepackage[abs]{overpic}
\begin{document}
	\author{Luis M. Canonico}
	\affiliation{Instituto de Física, Universidade Federal Fluminense, 24210-346 Niterói RJ, Brazil}
	\author{Tatiana G. Rappoport}
	\affiliation{Instituto de Física, Universidade Federal do Rio de Janeiro, Caixa
		Postal 68528, 21941-972 Rio de Janeiro RJ, Brazil}
	\author{R. B. Muniz}
	\affiliation{Instituto de Física, Universidade Federal Fluminense, 24210-346 Niterói RJ, Brazil}
	
	\title{Spin and Charge Transport of Multi-Orbital Quantum Spin Hall Insulators }	
	\begin{abstract}
		The fabrication of bismuthene on top of SiC paved the way for substrate engineering of room temperature quantum spin Hall insulators made of group V atoms.  We perform large-scale quantum transport calculations in these 2d materials to analyse the rich phenomenology that arises from the interplay between topology, disorder, valley and spin degrees of freedom. For this purpose, we consider a minimal multi-orbital real-space tight-binding Hamiltonian and use a Chebyshev polynomial expansion technique. We discuss how the quantum spin Hall states are affected by disorder, sublattice resolved potential and Rashba spin-orbit coupling.
	\end{abstract}
	
	\maketitle
	\graphicspath{{./figures/}}
	
	In recent years,  topological insulators have caught the attention of the condensed matter physics community, due to their exotic properties and potential applications. These system are characterized by a gapped bulk that hosts metallic spin-polarized edge states, which are topologically protected by time-reversal symmetry \cite{Colloquium-kane-Mele,Topological-insulators-analogy-w-superconductors}. In 2D topological insulators, also known as quantum spin Hall insulators (QSHI), edge states are extremely robust against disorder because the only available backscattering channel is forbidden by topology \cite{Kane-MeleZ2TopologicalInsulator, QuantumSpinHall-dissordereffects, QSHIdisorder2}.  Consequently, they are promising materials for the development of advanced electronic/spintronic devices \cite{SpintronicAllanMcdonald,Wolf-SpintronicDevices}.
	
	Despite being predicted more than a decade ago \cite{Kane-Mele-Graphene,QuantumWells-Bernevig}, the production of room temperature QSHI remains a challenge~\cite{Konig-ExperimentalHgTe,quantumwell-InAsGaSb, Germaneneandsilicene-TI,Silicene-germanene-stanene-TI,Bismuth-Localized-edgestates,stanene,antimonene}. In fact, only recently a QSHI state was observed in a truly 2D system~\cite{WTe,WTe2}. However, due to its small band gap, cryogenic temperatures are still required to unveil its attributes. 
	
	One possibility of obtaining QSHIs with relatively large band gaps that display a robust topological phase, is by exploring honeycomb lattice materials whose electronic states are reasonably well described by a Hamiltonian that involves basically the $p_x $ and $p_y$ orbitals \cite{grow-largegap-substrates,MultiorbitalHoneycomb-Wu}. Their energy bands exhibit a Dirac cone in the vicinity of the $K$ point, similar to that of graphene, however accompanied by two extra narrow bands symmetrically located at low and high energies, respectively~\cite{flat-bands-Wu}. The presence of a sizeable spin-orbit coupling (SOC) induces band gaps that are comparable to atomic level splittings\cite{MultiorbitalHoneycomb-Wu}, and the system behaves as a QSHI with three topological gaps. This model has been studied in the context of optical lattices \cite{flat-bands-Wu,px-py-WU,Anomaloushall-Optical-Wu}, where the $p_x$-$p_y$ Hamiltonian is achieved by orbital filtering processes, which removes the $p_z$ orbital from the vicinity of the Fermi level. Furthermore, in the limit of vanishing $\pi$-bonding, the narrow bands that accompany the Dirac cones in the energy spectrum become completely flat, favouring the appearance of strong correlation effects such as the Wigner crystallisation of spinless fermions \cite{flat-bands-Wu,px-py-WU}.\\
	
	Recently Reis \textit{et al.} reported the realisation of a condensed matter analogue of $p_x$-$p_y$ honeycomb systems in flat bismuthene\cite{ReisBismutheneExperimental}. They used a SiC substrate to grow a single planar layer of  bismuth, which is stabilised by the substrate that also acts as an orbital filtering platform. This approach may be regarded as a new paradigm to produce QSHIs with large gaps, where the orbital properties of the system may be tuned by a convenient choice of the substrate.\\
	
	Motivated by these findings, we have investigated the electronic and spin properties of two-dimensional $p_x$-$p_y$ honeycomb lattices. We perform numerical quantum transport calculations for this class of systems in presence of Anderson disorder and vacancies, and examine the robustness of the topological gaps against disorder and the presence of Rashba SOC.  Since the strengths of the $\sigma$- and $\pi$-bondings in these systems are material dependent, and are also affected by the substrate upon which it is deposited, we have also inspected the changes produced in their electronic structures by varying the ratio between the $\pi$- and $\sigma$-bonding strengths. In addition, we show that when the sublattice symmetry is broken by a weak potential, the system exhibits a strong spin-valley locking effect, similar to the one observed in transition-metal dichalcogenides.  
	
	
	The first principles calculations reported in Refs.~\onlinecite{ReisBismutheneExperimental,gao2018spin,zhou2018giant} show that the low-energy electronic properties of planar bismuthene (and of other elements of the VA-group), grown on a SiC substrate, or functionalized with halogen atoms, are reasonably well described by an effective tight-binding model Hamiltonian including only two orbitals ($p_x$ and $p_y$) per atom. Thus, to explore the transport properties of these novel 2D material we consider the model Hamiltonian $\mathcal{H} = H_{0} + H_{I} + H_{R}$, where
	
	\begin{align}
		H_0=&\sum_{\langle i j\rangle} \sum_{\mu \nu s} t_{i j}^{\mu \nu}{p^\dagger_{i \mu s}}p_{j \nu s}+ \sum_{i \mu s}\epsilon_{i} p^\dagger_{i \mu s} p_{i \mu s}
		\label{eqn:HNN}
	\end{align}
	
	\noindent represents the electronic kinetic energy plus a spin-independent local potential. Here, $i$ and $j$ denote the honeycomb lattice sites positioned at $\vec{R}_i$ and $\vec{R}_j$, respectively. The symbol $\langle i j \rangle$ indicates that the sum is restricted to the nearest neighbour (n.n) sites only. The operator $p^{\dagger}_{i \mu s}$ creates an electron of spin $s$ in the atomic orbital $p_\mu$ ($\mu=x,y$) centred at $\vec{R}_i$; $s=\,\uparrow,\downarrow$ labels the two electronic spin states, and $\epsilon_i$ is the atomic energy at site $i$. The transfer integrals $t_{i j}^{\mu \nu}$ between the $p$ orbitals centred on n.n atoms are parametrised according to the standard Slater-Koster tight-binding formalism \cite{Slater-Koster}. They depend on the direction cosines of the n.n. interatomic directions, and may be approximately expressed as linear combinations of two other integrals ($V_{pp\sigma}$ and  $V_{pp\pi}$) involving the $p_{\sigma}$ and $p_{\pi}$ orbitals, where $\sigma$ and $\pi$ refer to the usual components of the angular momentum around these axes.\\
	
	The second term $H_I$ describes the intrinsic atomic SOC that couples the $p_x$ and $p_y$ orbitals, and it is given by
	
	\begin{equation}
		H_I= i \lambda_I  \sum_{j s} \sigma^z_{s s}  p^\dagger_{j y s} p_{j x s} + H.c. \,,
	\end{equation}
	
	\noindent where $\lambda_{I}$ is the strength of the intrinsic SOC, and  $\sigma^z_{s s}$ are the diagonal elements of the usual Pauli matrix $\sigma^z$. \\

	The third term $H_R$ simulates a Rashba spin-orbit contribution, activated by the inversion symmetry breaking of the system caused by the presence of the substrate. 
	
	\begin{equation}
		H_{R} = 2i\lambda_{R}\sum_{\langle i,j \rangle}\sum_{\mu \nu s}
		p^\dagger_{i \mu \bar{s}} \left[\hat{z}\cdot\left(\vec{\sigma}\times\hat{e}_{ij}\right)\right]_{\bar{s} s}p_{j \nu s} + H.c.\,,
	\end{equation}
	
	\noindent where $\vec{\sigma}$ symbolises the Pauli vector, $\hat{e}_{ij}$ denotes the unit vector along the n.n. intersite direction of $\vec{R}_j - \vec{R}_i$,  $\lambda_{R}$ is the Rashba SOC constant, and $\bar{s}$ designates the opposite spin direction specified by $s$. $H_R$ is a multi-orbital generalisation of the Rashba term considered by Kane and Mele \cite{Kane-MeleZ2TopologicalInsulator,MultiOrbitalHgTehoneycomblattices,SpinOrbitInteractionInDorbitalNobleMetal}.\\
	
\begin{figure}[h]
		\centering
		\includegraphics[width=0.95\columnwidth]{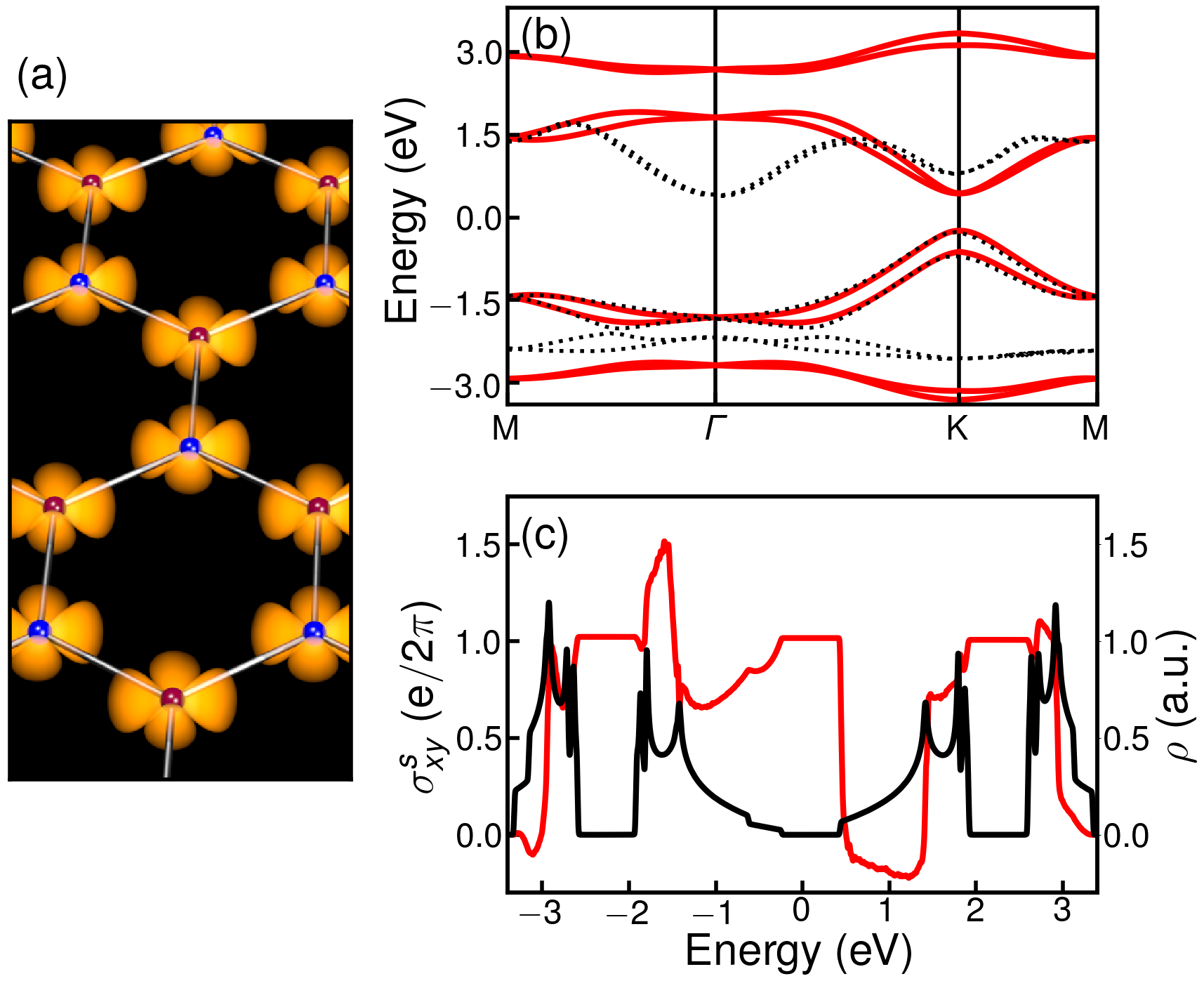}
		\caption{(a) Schematic representation of the $p_x$-$p_y$ orbitals in the honeycomb lattice. (b) Comparison between the first principles energy bands of planar bismuthene (black dotted line) and our effective $p_x$, $p_y$ tight-binding approximation (red solid line). (c) Numerical calculations of the spin Hall conductivity (red line) and of the density of states (black line).}\label{fig:bandas}
	\end{figure}

	We start by determining the parameters of this minimal model in order to fit the low-energy features around the chemical potential of the first-principles band-structure calculations for Bismuthene/SiC reported in Refs.~\onlinecite{ReisBismutheneExperimental,gao2018spin,zhou2018giant}. Panel (a) of figure \ref{fig:bandas} illustrates the $p_x$ and $p_y$ orbitals centred on the atomic sites of a honeycomb lattice that are involved in this simplified description. For the pristine case we choose the energy origin at $\epsilon_i = 0\,\forall i$, and our best fit is obtained with the parameters listed in table \ref{tab:parametrosSK}. 
	
	\begin{table}[h] 
		\centering
		
		\caption{Nearest-neighbour two-centre energy integrals, and spin orbit coupling constants (all in eV).}
		\label{tab:parametrosSK}
		\begin{tabular}{c c c}
			\hline	
			two-centre integrals  & Intrinsic SOC & Rashba SOC \\
			\hline
			$V_{pp\sigma}=+1.815$ & $\lambda_{I} = 0.435$ & $\lambda_{R} = 0.032$ \\
			$V_{pp\pi} = -0.315$ & {} & {}
			\\
		\end{tabular}
	\end{table}
	
	The values of our two-centre integrals differ slightly from those of Ref.~\onlinecite{li2018new} because in our procedure we emphasised an overall fitting of the upper valence bands calculated from first-principles  \cite{ReisBismutheneExperimental}. A comparison between our tight-binding bands and the latter is shown in panel (b) of the same figure, evincing that they are in fairly good agreement with each other. 
	
	This effective tight-binding model provides a reasonable description of the low-energy electronic properties of these systems, including their main topological characteristics. Its simplicity also makes it quite useful for carrying out large scale computational simulations in order to explore the quantum transport properties of these systems.	
	
	Here we shall make use of Chebyshev polynomial expansions \cite{SilverKPM,WeisseKPM} for this purpose. This real-space method has been employed with great success to analyse quantum transport properties of two dimensional systems \cite{Aires-JoaoKPM,KPM-DisorderGraphene,Graphene-Aires-Largescale, JoseTMDGraphene, Canonico2018}, by virtue of its high accuracy, stability and scalability. We calculate the longitudinal charge conductivity $\sigma_{xx}$ and the transverse spin-Hall conductivity $\sigma^s_{xy}$ with the help of the Kubo-Bastin formula\cite{Bastin-FormulaConductividad}, utilising an efficient implementation of the Chebyshev polynomial expansion, developed by J. H. Garc\'ia \textit{et al}\cite{Jose-TatianaKPM} and now included in the open-source software KITE~\cite{KITE} (further technical details can be found in the accompanying supplementary material).
	
	Results of our tight-binding calculations for the density of states $\rho$ and $\sigma^s_{xy}$ as functions of energy are shown in panel (c) of Fig. \ref{fig:bandas} for the planar Bi/SiC system. We notice that the three energy gaps depicted in the density of states curve coincide with quantised plateaus in the transverse spin Hall conductivity, indicating the materialisation of QSH states in these energy ranges. We remark that our real-space numerical calculations invariably include a large number ($\approx 10^6$) of lattice sites. 
	
	It is worth recalling that when $s_{z}$ is conserved, it is possible to classify the topological properties of the system in terms of invariant Chern integers $n_s$, associated with the two independent spin sectors. Time-reversal symmetry imposes that the total Chern integer $n = n_\uparrow + n_\downarrow = 0$, which leads to a quantised spin-Hall conductivity proportional to $2n_{\uparrow}$. As discussed in Ref.~\onlinecite{MultiorbitalHoneycomb-Wu}, without sublattice symmetry breaking, and for $0<\lambda_{I}<\frac{3}{2}V_{pp\sigma}$, our system is characterised by a band-Chern number sequence $(1,0,0,-1)$ associated with the $\uparrow$-spin sector, which exhibits a quantised spin-Hall conductivity. When the Rashba SOC is taken into account, $s_z$ no longer commutes with the Hamiltonian. Nevertheless, the system's topological features may be categorised either in terms of the spin-Chern number $\mathcal{C}_{s}$, or by the $\mathbb{Z}_2$ topological index \cite{HaldaneSpinCherNumbers,SpinChernNumbersProdanEmil,Kane-MeleZ2TopologicalInsulator,FukuiAlgorithm,FukuiKaneZ2}. Although the Rashba coupling tends to close the central gap, the topological gap sizes opened here by the intrinsic SOC are relatively large, and the effects of the RSOC are not sufficiently strong to change the overall topological order of the system, as our numerical calculations clearly show. Equally remarkable is the non-trivial topological behaviour of the bands in the energy ranges of the lateral gaps.
	
	\color{black} 
	
\begin{figure}[h]
		\centering
		\includegraphics[width=0.48\textwidth,trim={0cm, 0cm, 0cm, 2cm}]{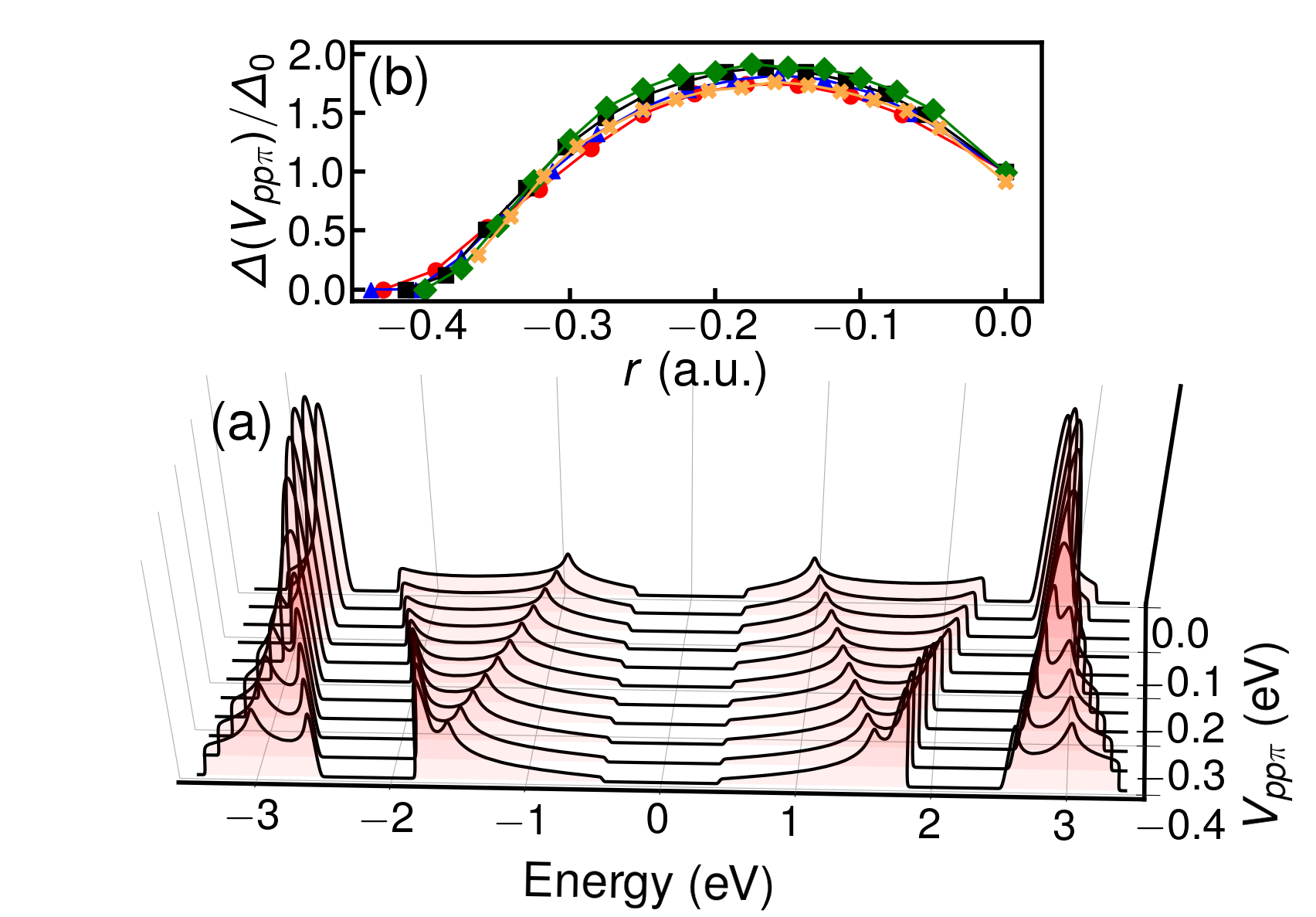}
		\caption{(a) Numerical calculation of the density of states of a $p_x,p_y$ honeycomb honeycomb system as function of $V_{pp\pi}$. (b) Variation of size of the lateral topological gap relative to $\Delta_0=\Delta(V_{pp\pi}=0)$, as function of the ratio $V_{pp\pi}/V_{pp\sigma}$, for: $V_{pp\sigma}=1.4\text{ eV}$ (red dots), $V_{pp\sigma}=1.6\text{ eV}$ (blue triangles), $V_{pp\sigma}=1.815\text{ eV}$ (black squares), $V_{pp\sigma}=2.0\text{ eV}$ (green diamonds) and $V_{pp\sigma}=2.2\text{ eV}$ (orange crosses).}\label{fig:figura2}	
	\end{figure}

	The role played by $V_{pp\pi}$ in the electronic properties of optical lattices is usually negligible due to perfect orbital confinement \cite{flat-bands-Wu,px-py-WU,Anomaloushall-Optical-Wu}. However, in condensed matter systems the relative strength between $V_{pp\pi}$ and $V_{pp\sigma}$ varies with the monolayer/substrate combination, and it may also be altered by suitable functionalizations. Therefore, it is instructive to examine the effects caused in the density of states (DOS) of our system when the ratio $r=V_{pp\pi}/V_{pp\sigma}$ is varied. Our results, calculated by means of a Chebyshev expansion for $\lambda_{R}=0$, are shown in Fig \ref{fig:figura2} (a). We notice that the central-gap size is basically dictated by the intensity of the intrinsic SOC~\cite{MultiorbitalHoneycomb-Wu,px-py-WU}, being virtually independent of $r$. Also, as $V_{pp\pi}$ becomes more negative, the bandwidths of the top and bottom bands increase, because the curvature of the energy bands grow, and so does the size of the lateral gaps. In panel (b) of Fig. \ref{fig:figura2}, we show results for the lateral-gap width $\Delta_L$, calculated as a function of $r$, for different values of $V_{pp\pi}$. They are expressed in units of $\Delta_L(0)$, which corresponds to the value of $\Delta_L$ obtained for $V_{pp\pi}=0$. The maximum value of the lateral gaps is reached for $r \approx  -0.18$, and then decreases to zero at $r\approx-0.4$.
	
	Our results suggest that a suitable choice of substrate can produce thermally robust quantum spin Hall insulators, and possibly used to manipulate the lateral topological gap sizes, as well as the dispersion of the flat bands in these systems, opening ways to explore the interplay between electronic correlation and topology.\\
	
	\color{black}
	
\begin{figure}[h]
		\centering
		\includegraphics[width=0.47\textwidth]{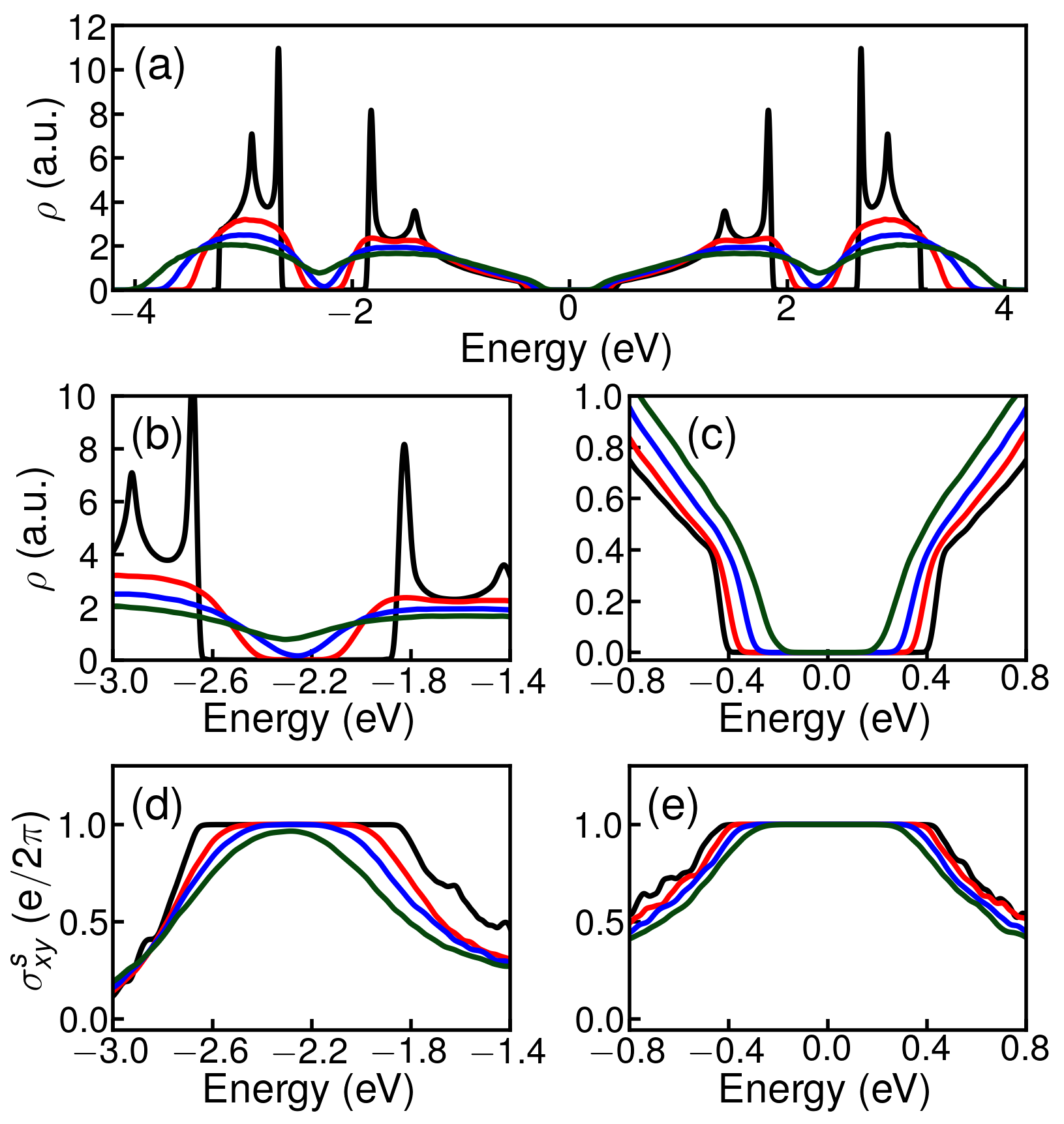}	
		\caption{Numerical calculation of the density of states [(a), (b) and (c)] and spin Hall conductivity [(d) and (e)], of a $p_x,p_y$ honeycomb system with Anderson disorder strengths: $W=0.05$ eV (black line), $W=0.8$ eV (red line), $W=1.2$ eV (blue line) and $W=1.6$ eV (green line).}\label{fig:EffectsAnderson}
	\end{figure}

	Two-dimensional materials are susceptible to imperfections of various types, which can strongly influence their transport properties. It is therefore instructive to simulate the presence of disorder in these systems and examine its consequences. To do so, we consider a disordered on-site potential $\epsilon_i$ that takes randomly distributed values between $[-W/2, W/2]$. In order to facilitate the comparison of how disorder affects the three topological gaps, we choose a value for $V_{pp\pi}$ that makes them approximately of the same size.  Our results for both $\rho$ and $\sigma^s_{xy}$, calculated as functions of energy, for $\lambda_{R}=0$, and different Anderson disorder strengths $W$, are depicted in Figure \ref{fig:EffectsAnderson}. Panel (a) shows that disorder broadens $\rho(E)$ as expected, but it clearly affects the energy gaps differently. To highlight the results we take a close-up of the lateral- and central-gap energy regions in panels (b) and (c) of the same figure, respectively. It is evident that the lateral gaps are less robust to effects of scalar disorder than the central one. The reason this happens lies in the distinct characteristics of the electronic states around those gaps. The localisation of massless Dirac fermions is qualitatively different from those governed by the Schr\"odinger equation~\cite{Mirlin}. Indeed, while the opening of the central gap around the $K$ and $K'$ points involves low-energy electronic states with linear energy dispersion relation, which are formally described by a massless Dirac equation, the lateral gaps in the vicinity of the $\Gamma$ point encompass approximately parabolic bands that represent Schr\"odinger-like electrons. 
	
	Panels  (d), and (e) illustrate the impact of disorder on the quantised spin-Hall conductivity plateaus associated with the topological gaps. The robustness of the central gap assures that the spin-Hall conductivity is fairly insensitive to weak Anderson disorder. In contrast, the size of the lateral gaps decreases more rapidly as localised states produced by the disorder begin to populate the inner region of this gap, and for $W\sim1.6\text{ eV}$ the quantum spin-Hall state is virtually destroyed. This class of materials thus exhibit two distinct regimes of disorder, which may be explored in a single sample of elements whose central and lateral gaps are accessible by gating. 
	
	We have also investigated how the presence of vacancies would affect the topological states of these systems. Our results show that they are fairly robust to this type of inhomogeneities. The plateaus in $\sigma^s_{xy}$ remain nearly unchanged for concentrations up to $1\%$ of vacancies, although it is noteworthy that in the neighbourhood of $E=0$ it deviates from the quantised value $e/2\pi$, due to the presence of impurity states in this energy range~\cite{disorderQSHI3}. Further results and analyses of the influences of Anderson and vacancy disorders, as well as the strength of the RSOC on the topological properties of these materials are presented in the SM.\\
	
	\begin{figure}[t]
		\centering
		\includegraphics[width=0.95\columnwidth]{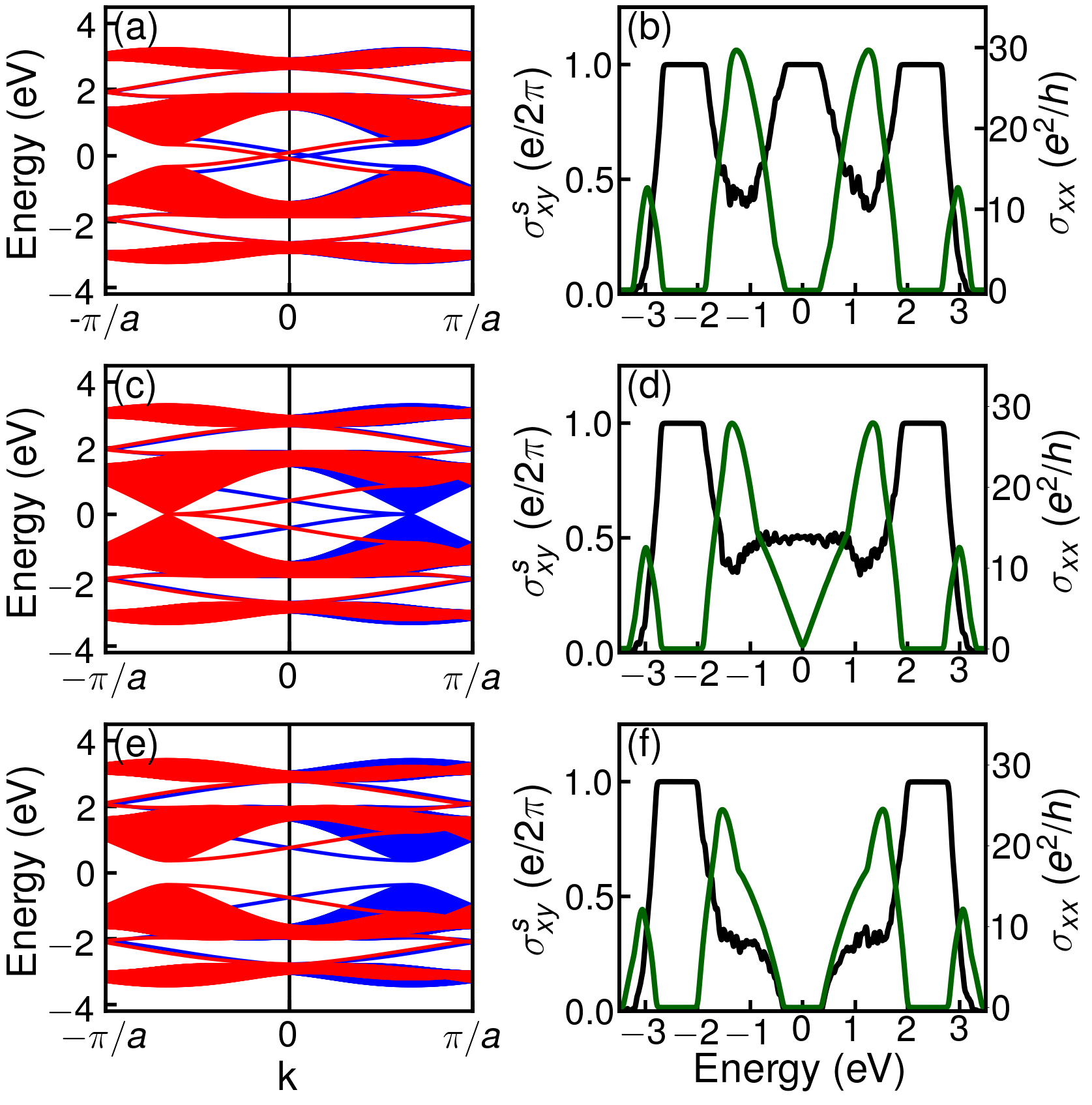}
		\caption{\textbf{Left panels}: Energy bands for a $30$ nm wide zigzag nanoribbon described by a $p_x$-$p_y$ Hamiltonian on a honeycomb lattice. The calculations were performed with $V_{pp\sigma}=1.815\text{ eV}$, $V_{pp\pi}=-0.315\text{ eV}$, $\lambda_{I}=0.435\text{ eV}$, $\lambda_{R}=0$ eV, for different values of the staggered potential: (a) $V_{AB}=0.1\text{ eV}$, (c) $V_{AB}=0.435\text{ eV}$ and (e) $V_{AB}=0.8\text{ eV}$. \textbf{Right panels}: spin-Hall conductivity (black line) and normalised longitudinal conductivity (green line)  calculated as functions of energy for $V_{pp\sigma}=1.815\text{ eV}$, $V_{pp\pi}=-0.315\text{ eV}$, $\lambda_{I}=0.435\text{ eV}$, $\lambda_{R}=0$ eV, $W=0.05\text{ eV}$, and different values of the stag gered potential: (b) $V_{AB}=0.1\text{ eV}$, (d) $V_{AB}=0.435\text{ eV}$ and (f) $V_{AB}=0.8\text{ eV}$.}\label{fig:figure4}\end{figure}
	
	Finally, in order to examine the interplay between spin and valley degrees of freedom, we consider an staggered on-site potential that breaks the inversion symmetry between the two interpenetrating triangular sublattices A and B. We assume that $\epsilon_i =+V_{AB}$ when $i$ denotes a site of sublattice A, and $\epsilon_i = -V_{AB}$ when it labels a site of the sublattice B. This can be achieved either by a suitable substrate choice~\cite{Theoretical-BilayerBismuth/SIC} or by selective functionalisation~\cite{zhou2018giant}. To inquire into the explicit presence and characteristics of edge states under those conditions, we begin by examining the band structure of a $30$ nm wide zigzag nanoribbon described by this simplified model.  The results calculated for different values of $V_{AB}$ are shown in the left panels of figure \ref{fig:figure4}. We notice that while $V_{AB} < \lambda_I$ the size of the central gap reduces as $V_{AB}$ increases, yielding edge states with opposite spin polarisations at the two inequivalent valleys. As a result, spin and valley are locked at the band edges~\cite{spin-valley}, but {\it the topological properties are preserved}, as shown in panel (b) of the same figure, which illustrates our bulk calculations for both $\sigma^s_{xy}$ and $\sigma_{xx}$ as functions of energy, employing the same set of parameters. This class of materials clearly display both spin- and valley-Hall effects, and for other elements of group V that exhibit a direct gap, as antimonene~\cite{antimonene,li2018new} for example, this spin-valley coupling may be explored for opto-spintronic applications~\cite{reviewspinvalley,spinvalley2,reviewspinvalley}. Panels (c) and (d) illustrate the special case in which $V_{AB} = \lambda_I$. In panel (c) we observe how the gap for one spin in this case closes in each cone, remaining opened for the opposite spin state. This is compatible with the fact that the central plateau in $\sigma^s_{xy}$ assumes the value $e/4\pi$, while $\sigma_{xx}$ displays a linear behaviour in this energy range, arising from states that are protected against spin-independent inter-valley scattering. Panels (e) and (f) depict the case where $V_{AB} > \lambda_I$. Here, the gaps re-open after a topological phase transition. The new state shows strong spin-valley coupling, but the edge states disappear as illustrated in panel (e). This explains the vanishing of the spin-Hall conductivity in the central gap, and its quantised presence in the energy ranges of the lateral ones. 
	
	In summary, we modelled a new series of substrate engineered 2D materials with a $p_x$-$p_y$ honeycomb Hamiltonian where the interaction between the substrate and the layer can tune the size of two of the three topological gaps of the system. We have performed large-scale real-space quantum transport calculations and show that while the lateral gaps are sensitive to Anderson disorder, the central gap is robust, verifying that the quantum spin-Hall phase survives even for strong disorder. The opposite occurs in the case of vacancies, and the difference can be understood in terms of the nature of the carriers in the vicinity of the gaps.  Finally, we demonstrate that these materials can exhibit a strong spin-valley coupling combined with quantum spin-Hall effect when the sublattice symmetry is broken.  Our quantum transport calculations show that this new class of multi-orbital 2d materials presents very robust topological phases and also spin-valley locking. Such characteristics, which are preserved in the presence of relatively strong disorder and high temperatures, are promising for the development of novel electronic and spintronic applications.  
	
	We would like to acknowledge CNPq/Brazil and INCT Nanocarbono for financial support and NACAD/UFRJ for high-performance computing facilities. 
	
	\pagebreak

	\pagebreak
	\begin{widetext}
		\begin{center}
			\textbf{\large Supplementary Material}
		\end{center}
	\end{widetext}
	\setcounter{equation}{0}
	\setcounter{figure}{0}
	\setcounter{section}{0}
	\setcounter{table}{0}
	\setcounter{page}{1}
	\renewcommand{\theequation}{S\arabic{equation}}
	\renewcommand{\thetable}{S\Roman{table}}
	\renewcommand{\thefigure}{S\arabic{figure}}

	\section{Chebyshev Polynomial Expansion}

	Our computational simulations were carried out by means of a Chebyshev polynomial method. The procedure requires a rescaling of the Hamiltonian to restrict its energy spectrum inside the interval $(-1,1)$ to guarantee the polynomial-series expansion convergence.  This is accomplished by defining $\mathcal{\tilde{H}} = \left(\mathcal{H}-b \right)/a$, whose eigenvalues are given by $\tilde{E} = \left(E-b \right)/a$.  Here $a=\left(E_{T}-E_{B}\right)/(2-\epsilon)$, and $b=\left(E_{T}+E_{B}\right)/2$; $E_{T}$ and $E_{B}$ represent the top and bottom energies of the spectrum, respectively, and $\epsilon$ is a small cut-off parameter introduced to prevent numerical instabilities.\\
	
	The expansion of the density of states $\rho$ in terms of the Chebyshev polynomials $T_{m}(\tilde{E})=\cos(m\arccos(\tilde{E}))$, up to order $N$, is given by 
	\begin{equation}
	\rho(\tilde{E}) = \frac{1}{\pi\sqrt{1-\tilde{E}^2}} \sum_{m=0}^{N-1}\mu_{m}g_{m}T_{m}(\tilde{E}),
	\end{equation}
	where $g_{m}$ is a kernel introduced to damp the Gibbs oscillations produced by truncation of the series \cite{WeisseKPM}. The coefficients of the expansion are given by $\mu_{m} = \langle T_{m}\left(\tilde{\mathcal{H}}\right)\rangle$, where $\langle \dots\rangle$ denotes a configurational average over disorder. In order to reduce the computational costs involved in the calculations of $\mu_{m}$, we employ a stochastic trace evaluation (STE) method\cite{WeisseKPM} to determine them. Instead of taking the full trace, they are obtained by  
	\begin{equation}
	\mu_{m} \approx \frac{1}{R}\left\langle\sum_{r =1}^{R} \langle \phi_{r}|T_{m}(\tilde{H})|\phi_{r}\rangle \right\rangle\,,
	\end{equation}
	where $|\phi_{r}\rangle$ are complex random vectors defined as $|\phi_{r}\rangle = D^{-1/2}\sum_{i=1}^{D}e^{i\theta_{i}}|i\rangle$. Here,  $\lbrace{|i\rangle\rbrace}_{i=1,...,D}$ denotes the original site and orbital basis set, $D$ is total number of state vectors used in the Hamiltonian representation matrix, and $\theta_{i}$ represent their phases. $R$ is the total number of random vectors that are used in the calculation. The convergence of STE scales with $1/\sqrt{DR}$.\\
	
	\section{Chebyshev Expansion of the Kubo-Bastin formula}
	
	In our transport simulations, we used an efficient implementation of the Chebyshev polynomial expansion developed by J. H. Garc\'ia \textit{et al.}\cite{Jose-TatianaKPM}, to calculate the components of the conductivity tensors by means of the Kubo-Bastin formula\cite{Bastin-FormulaConductividad}:
	\begin{flalign}
	&\sigma_{\alpha \beta}(\mu, T) = \frac{i\hbar}{\Omega}\int_{-\infty}^{+\infty}dE f(E; \mu, T) \nonumber\\ &\times Tr \langle j_{\alpha}\delta(E-\mathcal{H})j_{\beta}\frac{dG^{+}}{dE} - j_{\alpha}\frac{dG^{-}}{dE}j_{\beta}\delta(E-\mathcal{H})\rangle, 
	\label{KB}
	\end{flalign}
	where $\Omega$ represents the area of the two-dimensional sample, $f(E;\mu,T)$ is the usual Fermi-Dirac distribution function of energy $E$, chemical potential $\mu$ and temperature $T$. $G^+$, and $G^-$ symbolise the advanced and retarded one-electron Green functions, respectively. Here, $j_\alpha$ and $j_\beta$ may symbolise either the spin- or the charge-density-current operators, which are involved in the corresponding definitions of the charge and spin-Hall conductivity tensor components of interest. In the case of the longitudinal-charge conductivity $\sigma_{x,x}(\mu, T)$, both $j_\alpha$ and $j_\beta$ in  equation (\ref{KB}) represent the charge current-density operator $j_{x}\equiv\frac{ie}{\hbar}\left[x,\mathcal{H}\right]$. For the transverse spin conductivity $\sigma^s_{x,y}(\mu, T)$, $j_\alpha = j_x$, and $j_\beta$ is substituted by the spin-current-density operator given by $j_{y}^s\equiv\frac{1}{2}\left\lbrace \sigma_{z},v_{y}\right\rbrace$, where $\sigma_z$ is the usual Pauli matrix, and $v_y$ denotes the $y$-Cartesian component of the velocity operator.  In our numerical calculations employing equation (\ref{KB}), both the one-electron propagators and the spectral functions are expanded in terms of Chebyshev polynomials of first kind in $\tilde{H}$. It is worth noting that an optimised version of this implementation is now part of the open-source code KITE~\cite{KITE}. More details about the method can be found in Refs. \onlinecite{Jose-TatianaKPM,Jose-tatiana}.\\
	
	\section{Effects of the Rashba spin-orbit coupling}
\begin{figure}[h]
		\centering
		\includegraphics[width=0.47\textwidth]{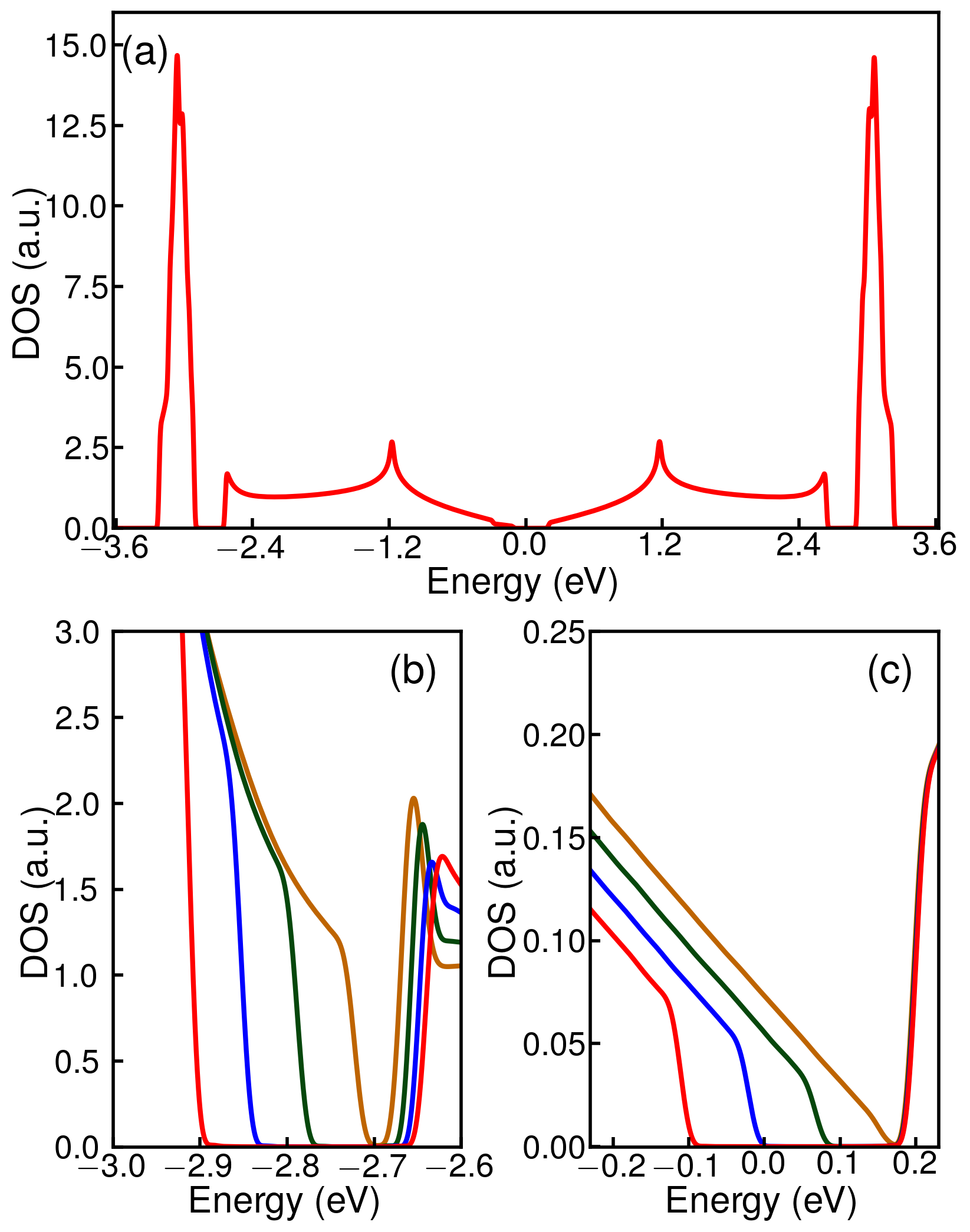}
		\caption{(a) Density of states of planar antimonene calculated with the parameters listed in table \ref{tab:parametrosSKantimonene}. (b) Close-up of the density of states of planar antimonene in the energy range of the lateral gap, calculated for different values of the Rasba SOC constant: $\lambda_{R}=0.015\text{ eV}$ (red line), $\lambda_{R}=0.030\text{ eV}$ (blue line), $\lambda_{R}=0.045\text{ eV}$ (green line), $\lambda_{R}=0.060\text{ eV}$ (brown line). (c) Close-up of the density of states of planar antimonene in the energy range of the central gap, calculated for the same values of $\lambda_R$, listed previously.}\label{fig:sm2}
	\end{figure}

	In this section, we explore the effects of the Rashba spin-orbit coupling (RSOC) in systems described by the $p_x$-$p_y$ simplified tight-binding model. Although planar antimonene/SiC has not been synthesised yet, we choose it as an example, because the atomic SOC in Sb is smaller than in Bi. Thus, its topological gaps are narrower, which highlights the impacts of the RSOC. In our tight-binding calculations we use the parameters listed in table \ref{tab:parametrosSKantimonene}, which were obtained by a fitting of first-principles band-structure calculations of planar antimonene/SiC\cite{li2018new}. The results for the density of states (DOS) of planar antimonene are depicted in panel (a) of figure \ref{fig:sm2}, Both the central and lateral gaps are clearly narrower in comparison with the corresponding ones for Bi (see panel (c)  in figure \ref{fig:bandas}), as expected. The influence of the RSOC on the topological gaps are evidenced in panels (b) and (c) of the same figure, which exhibit close-ups of the DOS in the energy ranges of the lateral and central gaps, respectively, calculated for different values of the RSOC constant. We note that the effect of the RSOC is more pronounced in the central gap than in the lateral ones.This difference is due to the character of the electronic states involved in the gap opening processes in these two energy regions of the spectrum. While RSOC produces a band splitting at the Dirac points for relativistic carriers, it is inactive at $k=0$ for Schr\"{o}dinger electrons.\\
	\begin{table}[h] 
		\centering
		\caption{Nearest-neighbour two-centre energy integrals, and spin orbit coupling constants (all in eV) of the $px,py$ planar antimonene.}
		\label{tab:parametrosSKantimonene}
		\begin{tabular}{c c c}
			\hline	
			Two-centre integrals  & Intrinsic SOC & Rashba SOC \\
			\hline
			$V_{pp\sigma}=+2.0$ & $\lambda_{I} = +0.2$ & $\lambda_{R} = +0.015$ \\
			$V_{pp\pi} = -0.11$ & {} & {}\\
		\end{tabular}
	\end{table}
	
	\section{Effects of disorder in the quantum transport properties of planar bismuthene}
	
\begin{figure}[h]
		\centering
		\includegraphics[width=0.47\textwidth]{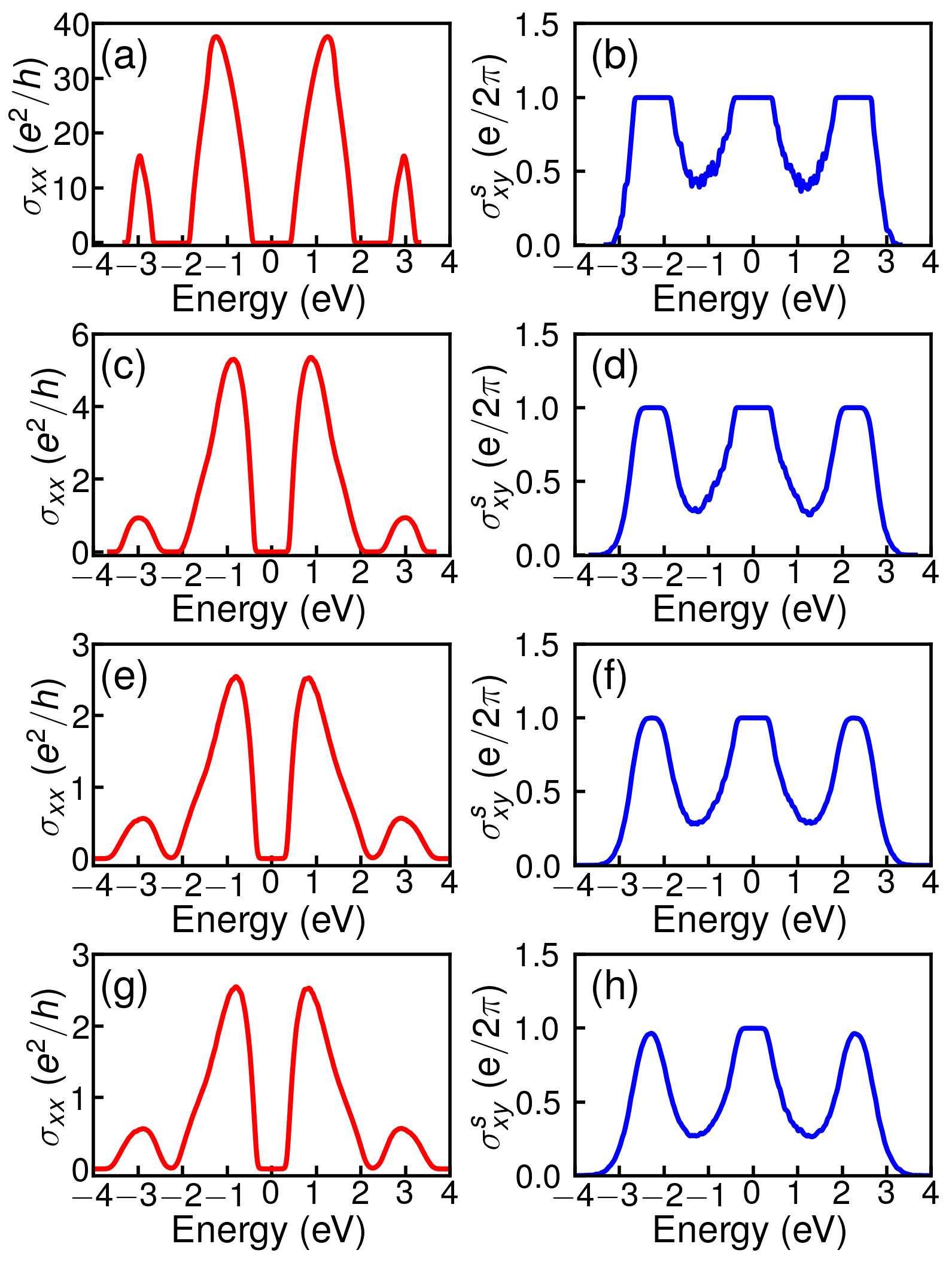}
		\caption{\textbf{Left panels:} Longitudinal charge conductivities of planar bismuthene calculated for different strengths $W$ of Anderson disorder: (a) $W=0.05$ eV, (c) $W=0.8$ eV, (e) $W=1.2$ eV and (g) $W=1.6$ eV. \textbf{Right panels:} Transverse spin conductivity of planar bismuthene calculated for different strengths $W$ of Anderson disorder: (b) $W=0.05$ eV, (d) $W=0.8$ eV, (f) $W=1.2$ eV and (g) $W=1.6$ eV.}\label{fig:sm3}
	\end{figure}

	In this section we present some supplementary results that illustrate the effects of both Anderson and vacancy types of disorder in the quantum transport properties planar bismuthene. Figure \ref{fig:sm3} shows the longitudinal-charge, as well as the transverse-spin conductivities calculated as functions of energy for different Anderson disorder strengths $W$. In the right column, we clearly observe that the longitudinal-charge conductivity significantly reduces when the Anderson-disorder strength increases, due to localisation effects, as expected. In contrast, the transverse spin conductivities, displayed in the right panels of the same figure, are much less affected, except for the lateral plateaus, which are much more susceptible to Anderson disorder than the central one. 
	
\begin{figure}[h]
		\centering
		\includegraphics[width=0.47\textwidth]{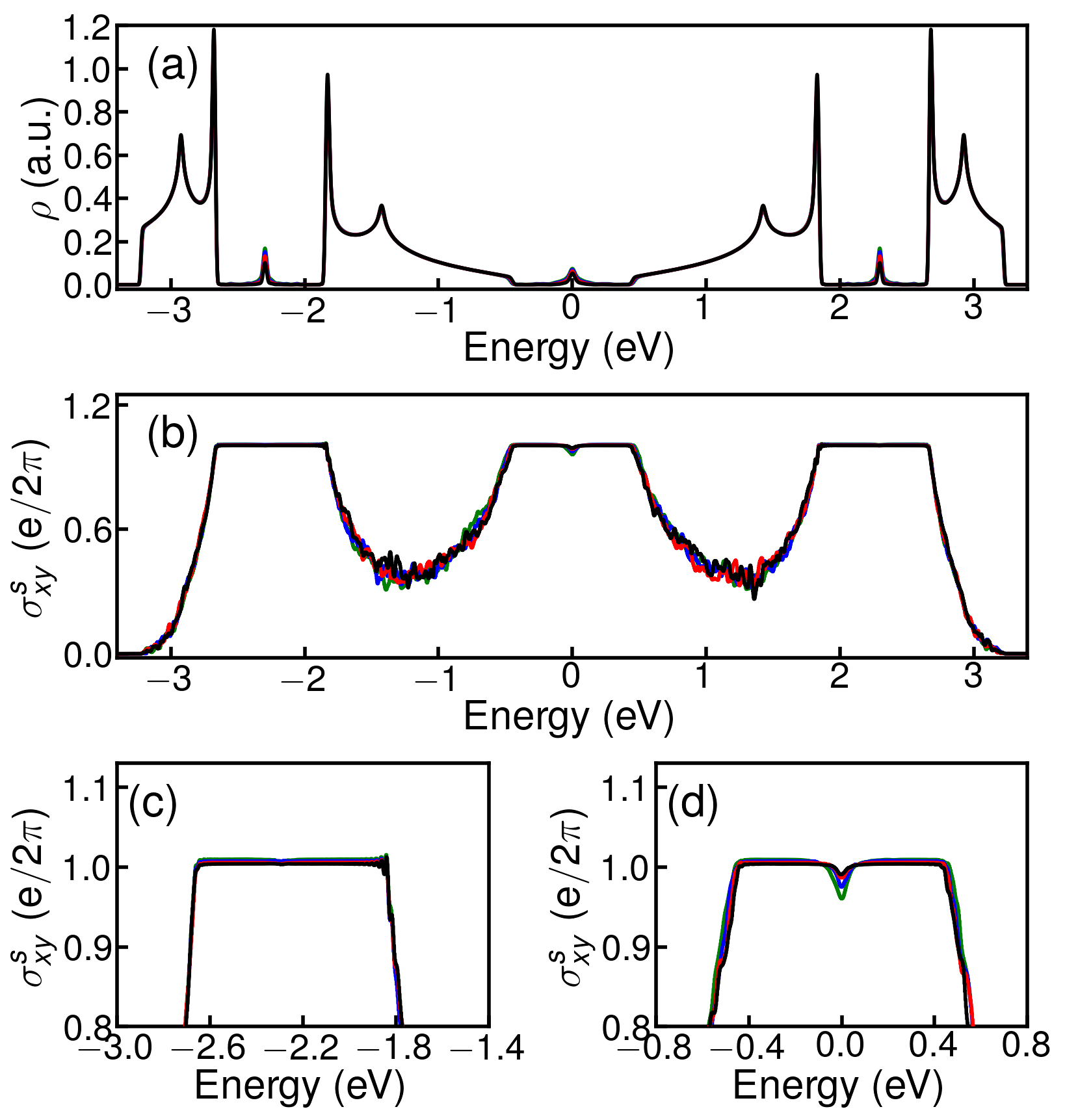}
		\caption{Density of states (a) and spin-Hall conductivities (b, c and d) of planar bismuthene calculated for different concentration values ($x$) of disordered vacancies: $x=0.4\%$ (black line), $x=0.6\%$ (red line), $x=0.8\%$ (blue line), $x=1.0\%$ (green line). }\label{fig:sm4}
	\end{figure}

	In figure \ref{fig:sm4}, we show the density of states and the transverse spin conductivities, calculated for planar bismuthene, as functions of energy, for different concentrations ($x$) of disordered vacancies. In panel (a) we notice the appearance of resonance peaks inside the three topological gaps, with increasing heights as $x$ increases. Results for the corresponding spin-Hall conductivities are exhibited in panels (b), (c) and (d) of the same figure. In panel (b) it is clear that overall form of the spin-Hall conductivity is little affected by the presence of vacancies, except in neighbourhood of $E=0$, where the its value becomes slightly lower than $e/2\pi$ that characterises the QSHI state. In panels (c) and (d) we take close-ups of the lateral and central plateaus displayed by the spin Hall conductivities in order to highlight these features. The resonance peaks that appear inside the lateral gaps of the DOS correspond to localised states that do not affect QSHI state. In contrast, the ones that emerge in the central gap, close to $E=0$, are associated with non-localised resonant states that are protected by chiral symmetry\cite{AiresCriticalDelocalizatoin} and reduce QSHI plateau in this energy range.\\
	
	\section{Numerical Implementation Details}
	
	In our quantum transport simulations for $\lambda_{R}=0$, we have used $D=4\times 512\times 512$, $N=1024$, and $R=300$ random vectors to perform the stochastic trace calculations. We choose a fixed Anderson disorder strength$W=0.1$eV and a temperature $T=100K$.\\
	
	For $\lambda_{R}\neq0$ we take $D=4\times 512\times 512$, $N=2560$, and $R=300$. We choose $W=0.1$ eV and a temperature $T=1\times10^{-5}K$ to reduce to the minimum the smearing of the conductivity plateaus.\\
	
	In our density of states calculations we used $D=4\times 1024\times1024$ sites, $N=2048$,  $R=300$, and $W=0.05$ eV.\\
	
\end{document}